\documentstyle[twocolumn,aps,psfig]{revtex}

\input{psfig}
\def\be{\begin{equation}}
\def\ee{\end{equation}}
\def\bea{\begin{eqnarray}}
\def\eea{\end{eqnarray}}
\begin{document}

\title{Strangeness enhancement from strong color fields at RHIC}

\author{M.~Bleicher$^{a,\xi}$\thanks{Feodor Lynen Fellow of the Alexander v. 
Humboldt Foundation, E -mail: bleicher@nta2.lbl.gov}, W. Greiner${}^{b}$, 
H.~St\"ocker${}^{b}$, N. Xu${}^{a}$}

\address{${}^a$ Nuclear Science Division,
Lawrence Berkeley Laboratory, Berkeley, CA 94720, U.S.A.\\}

\address{${}^b$Institut f\"ur
Theoretische Physik,  Goethe-Universit\"at,
60054 Frankfurt am Main, Germany}


\maketitle
\begin{abstract}

In ultra-relativistic heavy ion collisions, early stage 
multiple scatterings may lead to an increase of the color
electric field strength. Consequently, particle production -
especially heavy quark (and di-quark) production - is greatly
enhanced according to the Schwinger mechanism. We test this idea via the
Ultra-relativistic Quantum Molecular Dynamics model (UrQMD) for Au+Au
collisions at the full RHIC energy ($\sqrt{s} = 200$ AGeV).  Relative
to p+p collisions, a factor of 60, 20 and 7 enhancement
respectively, for $\Omega$ ($sss$), $\Xi$ ($ss$), and $\Lambda$,
$\Sigma$ ($s$) is predicted for a model with increased color
electric field strength.

\vfill 
\vspace{0.2in}

\noindent
\underline{LBNL-preprint: LBNL-46167}
\end{abstract}
\vspace{0.5in}

%--=========================================================================

 One of the major goals of the relativistic heavy ion collider (RHIC)
 at Brookhaven National Laboratory is to explore the phase diagram of
 hot and dense matter near the quark gluon plasma (QGP) phase
 transition.  The QGP is a state in which the individual hadrons
 dissolve into a gas of free (or almost free) quarks and gluons in
 strongly compressed and hot matter (for recent reviews on the topic,
 we refer to \cite{qgprev,mu1}).  The achievable energy- and baryon
 densities depend on the extend to which the nuclei are
 stopped during penetration, thus on centrality and
 bombarding energy.

 Strange particles, especially  multi-strange baryons (which have more than one
 strange quark) carry vital information about the collision dynamics
\cite{raf8286,koch86,koch88,senger99,JPG,stock99,rafelski96,geiss98}.
Relative strangeness abundancies have been proposed as a 
powerful tool for searching the transition 
 from hadronic matter to partonic matter in high energy nuclear
 collisions \cite{raf8286,koch86,koch88}. Indeed, strangeness 
 enhancement has be observed in heavy ion collisions at 
 all collision energies with different colliding systems. 

Recently, measurements by the WA97 and the NA49 collaborations 
clearly demonstrated the relative enhancement 
of the (anti-)hyperon yields ($\Lambda$, $\Xi$, $\Omega$) 
in Pb-Pb collisions as  
compared to p-Pb collisions \cite{and98a,and98,and99b,mar99,gab99,evans99}. 
The observed enhancement increases with the strangeness content ($|S|=1,2,3$)
of  the probe under investigation 
\cite{and98a,and98,and99b,mar99,gab99,evans99}. 
For the ($\Omega^- + \overline{\Omega^-}$)-yield the enhancement factor 
is as large as 15.

A number of different mechanisms are under debate to understand this 
strong increase in strangeness production with centrality and beam energy: 
\begin{itemize}
\item Equilibrated (gluon rich) plasma phase: 
Chemical and flavour equilibration times are predicted to be
shorter in a plasma phase than in a thermally equilibrated hadronic 
fireball of $T\sim 160\,$MeV \cite{raf8286}. Thus, a dominant
production mechanism for strangeness in an equilibrated gluon rich 
plasma phase (Hot glue scenario \cite{hot-glue}), might be the production 
of $s\overline{s}$ pairs via gluon fusion ($gg \rightarrow s \overline{s}$)
\cite{raf8286}. 
This might allow for strangeness equilibration within the 
lifetime of the QGP, resulting in strong strangeness enhancement 
compared to hadronic scenarios.

\item Baryon-junctions: A baryon junction exchange mechanism was
proposed to explain valence baryon number transport in nuclear
collisions. Recently it was extended to study midrapidity
(anti-)hyperon production \cite{Vance:1999pr}.  It was found that 
Baryon junction-anti-junction (J anti-J) loops can indeed enhance
anti-Lambda, anti-Xi, anti-Omega yields at SPS energies.  
While this mechanism leads to a reasonable description of central interactions,
it can not explain the enhancement
in the measured hyperon to anti-hyperon ratios with increasing centrality.

\item Diquark breaking and sea-diquarks:
A new version of the dual parton model, featuring an improved 
diquark breaking mechanism has been applied to explain the observed
strangeness enhancement at the full SPS energy \cite{Capella:1999cz}.
Here, the strange- and anti-baryon production invokes strings originating
from diquark-antidiquark pairs in the nucleon sea to reproduces the
observed yields of p and Lambda and their antiparticles. However
cascades are underestimated by 50\% and Omega's are underestimated by
a factor five.  Agreement with measured data can only be achieved in
this model if additional final state interactions among hadrons are
taken into account.

\item Strong color fields: Another kind of
string-hadronic models employs the Schwinger mechanism
\cite{schwing51} of a fragmenting color field (string) directly.
However, in central high energy heavy ion collisions the string
density can be so high that the color flux tubes overlap
\cite{biro84,sor92}.  Consequently, the superposition of the color
electric fields yields an enhanced particle production
\cite{biro84,sor92}.  In particular, the heavy flavors and diquarks
are dramatically enhanced due to a higher effective string tension
\cite{sor92,gyulassy90,gerland95}. 
Note that the strong color
field leads to the enhancement not only of the heavy flavors and diquarks 
but also enhances the high $p_t$ tail of the transverse momentum distribution 
of all created particles. 
However, the changes in the final $p_t$ are rather moderate in AA collisions 
due to final state interactions \cite{sor92}.
A consistent enhancement of hyperons and anti-protons is necessary to 
support this scenario. 
In fact, it has been shown  that this
approach is consistent with the measured anti-proton and hyperon
yields at SPS energies, if rescattering effects are taken into account
\cite{Soff:1999et,Bleicher:2000gj}.

\end{itemize}

 As a tool for our
 investigation of heavy ion reactions at RHIC the Ultra-relativistic
 Quantum Molecular Dynamics model (UrQMD 1.2) is applied \cite{urqmd}.
 UrQMD is a microscopic transport approach based on the 
 covariant propagation of
 constituent quarks and diquarks accompanied by mesonic and baryonic
 degrees of freedom.  It simulates multiple interactions of ingoing
 and newly produced particles, the excitation and fragmentation of
 color strings and the formation and decay of hadronic resonances.  At
 RHIC energies, the treatment of subhadronic degrees of freedom is of
 major importance.  In the UrQMD model, these degrees of freedom enter
 via the introduction of a formation time for hadrons produced in the
 fragmentation of strings
 \cite{andersson87a,andersson87b,sjoestrand94a}.  The leading hadrons
 of the fragmenting strings contain the valence-quarks of the original
 excited hadron. In UrQMD they are allowed to interact even during
 their formation time, with a reduced cross section defined by the
 additive quark model, thus accounting for the original valence quarks
 contained in that hadron \cite{urqmd}. Those leading hadrons
 therefore represent a simplified picture of the leading (di)quarks of
 the fragmenting string.  Newly produced (di)quarks do, in the present
 model, not interact until they have coalesced into hadrons --
 however, they contribute to the energy density of the system.  
 For further details about the UrQMD model,
 the reader is referred to Ref. \cite{urqmd}.

In the following two different scenarios
will be explored in order to study the speculation of a string tension increase
in heavy ion collision at RHIC energies: UrQMD calculations with the 
standard color flux tube break-up mechanism (i.e. a string tension 
$\kappa =1$~GeV/fm)
will be contrasted by UrQMD simulations with an in-medium $\kappa$
increased to 3~GeV/fm. According to the implemented Schwinger mechanism for the
string fragmentation ($m$ denoting the ((di-)quark masses)
\begin{equation}
\gamma_{X}=\frac{P(X\overline{X})}{P(q\bar{q})}=\exp
\left(- \frac{\pi
(m_{X}^2-m_q^2)}{\kappa}\right)\quad,
\label{gamm}
\end{equation}
this results in an enhancement of the strangeness and diquark production
probabilities to $\gamma_s=0.72$
and $\gamma_{qq}=0.46$ (for $\kappa=3$~GeV/fm) compared to $\gamma_s=0.37$
and $\gamma_{qq}=0.093$ for $\kappa=1$~GeV/fm. Note that a decrease of the
(di-)quark masses due to chiral symmetry restoration may lead to a similar
enhancement as the $\kappa=3$~GeV/fm scenario.

\begin{figure}[hb]
\centerline{\psfig{figure=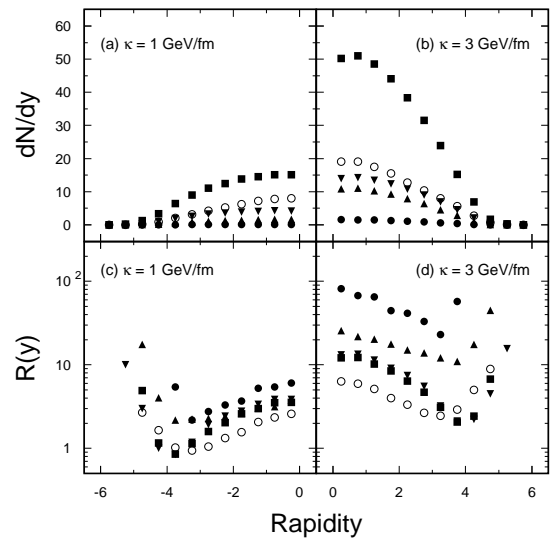,width=3.5in}}
\caption{Particle rapidity distributions and ratios from Au+Au (b = 2
fm) and p+p collisions at the RHIC full energy $\sqrt{s} = 200$AGeV.
(a) Rapidity distributions with string energy density $\kappa =
1$GeV/fm$^3$; (b) same as (a) with $\kappa = 3$GeV/fm$^3$; (c) ratio
$R(y)=\frac{\frac{dN}{dy}^{\rm Au}(y)/A_{\rm part}}{\frac{dN}{dy}^{\rm pp}(y)/2}$ 
of the particle yields per participant from central nuclear collisions
over that of the p+p collisions with $\kappa = 1$GeV/fm; (d) same
as (c) with $\kappa = 3$GeV/fm.
$\Omega^-+\overline{\Omega^-}$  (full circles), 
$\Xi^-+\overline{\Xi^-}$ (full triangles), $\Sigma^0+\overline{\Sigma^0}$
(reversed triangles), $\Lambda^0+\overline{\Lambda^0}$ (full squares) and
anti-protons (open circles).
\label{rapidity}}
\end{figure}

Fig. \ref{rapidity} depicts the particle rapidity distributions (a, b)
and ratios from Au+Au (b = 2~fm) to p+p collisions (c, d) at the RHIC
full energy $\sqrt{s} = 200$AGeV. The symbols denote:
$\Omega^-+\overline{\Omega^-}$  (full circles), 
$\Xi^-+\overline{\Xi^-}$ (full triangles), $\Sigma^0+\overline{\Sigma^0}$
(reversed triangle), $\Lambda^0+\overline{\Lambda^0}$ (full square) and
anti-protons (open circles). 
(a) shows the rapidity distributions with string tension
$\kappa = 1$GeV/fm; (b) shows the same as (a) with $\kappa = 3$GeV/fm.
(c) depicts the ratio $R(y)=\frac{\frac{dN}{dy}^{\rm
Au}(y)/A_{\rm part}}{\frac{dN}{dy}^{\rm pp}(y)/2}$  of the
particle yields per participant from central AA collisions over
that of the p+p collisions with $\kappa = 1$GeV/fm; (d) same as
(c), however with $\kappa = 3$GeV/fm for Au+Au.
%--============================================================================
%\newpage
%--============================================================================

A strong enhancement of the (anti-)hyperon production compared to
scaled pp interactions is predicted (cf. (c)). This enhancement is
purely due to rescattering between constituent (di-)quarks and
hadrons in the medium. If strong color fields, similar to Pb+Pb at
SPS \cite{Soff:1999et,Bleicher:2000gj}, are present, the 
model predicts a dramatic enhancement in the
multi-strange hadron production, up to a factor of 100 (in case of the
Omega) at midrapidity.

\begin{figure}[ht]
\vskip 0mm
%\vspace{-1.8cm}
\centerline{\psfig{figure=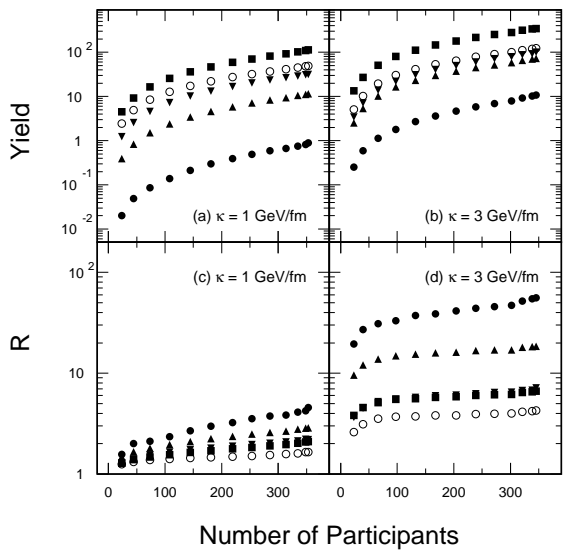,width=3.5in}}
\vskip 0mm
\vspace{-.0cm}
\caption{Particle 4$\pi$ yields and ratios as a function of number of
participants from Au+Au  and inelastic p+p collisions at the RHIC full
energy $\sqrt{s} = 200$AGeV. (a) yields with string energy density
$\kappa = 1$GeV/fm$^3$; (b) same as (a) with $\kappa = 3$GeV/fm$^3$;
(c) ratio $R=\frac{N^{\rm Au}(A_{\rm part})/A_{\rm part}}{N^{\rm pp}/2}$ 
of the particle yields per participant from central AA 
collisions over that of the p+p collisions with $\kappa =
1$GeV/fm$^3$; (d) same as (c) with $\kappa = 3$GeV/fm$^3$ for the Au+Au
interactions. $\Omega^-+\overline{\Omega^-}$  (full circles),  
$\Xi^-+\overline{\Xi^-}$ (full triangles), $\Sigma^0+\overline{\Sigma^0}$
(reversed triangles), $\Lambda^0+\overline{\Lambda^0}$ (full squares) and
anti-protons (open circles).
\label{participant}}
\end{figure}

The enhancement of the hyperon yields is strongly dependent on the
centrality of the events as shown in Fig. \ref{participant}: (a) shows the
particle yields with string tension $\kappa = 1$GeV/fm, while (b) 
is same as (a) with $\kappa = 3$GeV/fm. Fig. \ref{participant}(c) shows the
ratio $R=\frac{N^{\rm Au}(A_{\rm part})/A_{\rm part}}{N^{\rm pp}/2}$
of the particle yields per participant\footnote{The number of 
participating nucleons in this paper is defined as:\\
$A_{\rm part}= A_1+A_2 - \Sigma \, (\mbox{Nucleons with } p_T\leq 270$~MeV).
This prescription yields a reasonable parametrization of the experimental data
on $A_{\rm part}$.}
 from central AA collisions
over that of the inelastic p+p collisions with $\kappa = 1$GeV/fm.
(d) depicts the same as (c) 
with $\kappa = 3$GeV/fm for Au+Au
collisions. The symbols are: $\Omega^-+\overline{\Omega^-}$  (full circles),  
$\Xi^-+\overline{\Xi^-}$ (full triangles), $\Sigma^0+\overline{\Sigma^0}$
(reversed triangle), $\Lambda^0+\overline{\Lambda^0}$ (full square) and
anti-protons (open circles).
%=======================================================================
%\newpage
%======================================================================
 The 4$\pi$ yields and ratios as a function of number of participants from
Au+Au at the RHIC full energy $\sqrt{s} = 200$AGeV increase rapidly
with centrality (a, b) by two orders of magnitude when going from
peripheral ($A_{part} \approx 25$) to central collision.  The
enhancement itself is most pronounced when scaled with the number of
participating nucleons $A_{part}$ and compared to p+p collisions at
the RHIC full energy (c, d).

\begin{figure}[ht]
\vskip 0mm
%\vspace{-1.0cm}
\centerline{\psfig{figure=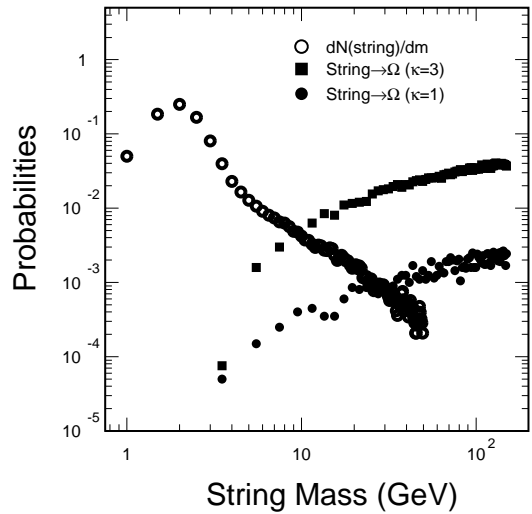,width=3.5in}}
\vskip 2mm
%\vspace{-1.0cm}
\caption{Probability distributions of strings (open circles). Filled
circles and squares represent the probabilities of strings to decay into
$\Omega$s with $\kappa$ = 1 and 3 GeV/fm, respectively. 
\label{probabity}}
\end{figure}

Finally, Fig. \ref{probabity} depicts an inside view into a central (b=2~fm)
Au+Au collision at the full RHIC energy. Open circles show the mass 
distribution of excited strings.  Filled circles and
squares represent the probabilities of a $[3]-[\,\overline 3\,]$ string of a
given mass to decay into $\Omega$s with $\kappa$ = 1 and 3 GeV/fm,
respectively.  The increase of the string tension
to $\kappa=3$~GeV/fm leads to a strong increase of the Omega
production over all inspected string masses (compare full circle with
full squares). However, the probability to excite a string with a large
mass is strongly suppressed, as can been seen by the exponential
decrease of probabilities above excitation energies of 2~GeV. Therefore,
most of the $\Omega$'s are produced from rather low mass
strings. This is in contrast to the expectations that early (high
energetic) collisions lead to high mass excitations and  give the
major contribution to the multi-strange hyperon yield.
%--============================================================================
%\newpage

%--============================================================================

 In conclusion, the UrQMD model has been applied to Au+Au reactions at
 RHIC energies. This model treats the dynamics of the hot and dense
 system by constituent (di-)quark and hadronic degrees of freedom.
 (Anti-)Hyperon yields in central Au+Au collisions at the full RHIC
 energy ($\sqrt s=200$~AGeV) strongly enhanced in comparison to
 inelastic pp interactions at $\sqrt s=200$~GeV or peripheral Au+Au
 collisions.  This enhancement grows dramatically with the strangeness
 content of the hyperon.  The present model predicts that strangeness
 enhancement occurs as a threshold effect already at rather small
 number of participants $(\approx 25)$ due to rescattering.
 Increasing the string tension, similar to SPS energies (or reducing
 the effective masses of the constituent quarks to the current quark
 mass values yields large additional enhancement, which grows with the
 strangeness content even stronger: $\Lambda$'s are enhancement by a
 factor of 7, while $\Omega$'s are enhanced by a factor of 60 compared
 to pp.  

\section*{Acknowledgements}

This research used resources of the National Energy Research
Scientific Computing Center (NERSC).  This work is supported by the
U.S. Department of Energy under contract No. DE-AC03-76SF00098, the
BMBF, GSI, DFG and Graduiertenkolleg 'Theoretische und experimentelle
Schwerionenphysik'.  M. Bleicher is supported by the A. v. Humboldt
foundation. M. Bleicher wants to thank the Nuclear
Theory Group at LBNL for financial support and fruitful discussions.

%--============================================================================

\end{document}